\documentclass{jpp}
\usepackage{graphicx}
\usepackage{epstopdf, epsfig}

\shorttitle{Radially anisotropic systems with $r^{-\alpha}$ forces}
\shortauthor{P.F. Di Cintio, L. Ciotti and C. Nipoti}

\title{Radially anisotropic systems with $r^{-\alpha}$ forces: equilibrium states}

\author{PIERFRANCESCO DI CINTIO\aff{1,2}
  \corresp{\email{pierfrancesco.dicintio@unifi.it}},
  L. CIOTTI\aff{3},
  \and C. NIPOTI\aff{3}}

\affiliation{\aff{1}Dipartimento di Fisica e Astronomia, Universit\`a di Firenze e Centro Studi Dinamiche Complesse, via Sansone 1 I-50022 Sesto Fiorentino, Italy
\aff{2}INFN - Sezione di Firenze
\aff{3}Dipartimento di Fisica e Astronomia, Universit\`a di Bologna, viale Berti-Pichat 6/2, I-40127 Bologna, Italy}

\begin{document}

\maketitle

\begin{abstract}
We continue the study of collisionless systems governed by additive $r^{-\alpha}$ interparticle forces by focusing on the influence of the force exponent $\alpha$ on radial orbital anisotropy. In this preparatory work we construct the radially anisotropic  Osipkov-Merritt  phase-space distribution functions for self-consistent spherical Hernquist models with $r^{-\alpha}$ forces and $1\leq\alpha<3$. The resulting systems are isotropic at the center and increasingly dominated by radial orbits at radii larger than the anisotropy radius $r_a$. For radially anisotropic models we determine the minimum value of the anisotropy radius $r_{ac}$ as a function of $\alpha$ for phase-space consistency (such that the phase-space distribution function is nowhere negative for $r_a\geq r_{ac}$). We find that $r_{ac}$ decreases for decreasing $\alpha$, and that the amount of kinetic energy that can be stored in the radial direction relative to that stored in the tangential directions for marginally consistent models increases for decreasing $\alpha$. In particular, we find that isotropic systems are consistent in the explored range of $\alpha$. By means of direct $N$-body simulations we finally verify that the isotropic systems are also stable. 
\end{abstract}
\section{Introduction}
It is well known that spherically symmetric, self-gravitating collisionless equilibrium systems in which a significant fraction of the kinetic energy is stored in low angular momentum orbits 
are dynamically unstable. The associated instability is known as Radial Orbit Instability (hereafter ROI, see e.g. \citealt{fridman84,bert01,bt2008}; for a recent discussion see also \citealt{poly15} and references therein). Usually, the amount of radial anisotropy in a spherical system is quantified by introducing the Fridman-Polyachenko-Shukhman parameter
\begin{equation}
 \xi\equiv\frac{2K_r}{K_t},
\end{equation}
where the radial and tangential kinetic energies are given respectively by
\begin{equation}
K_r=2\pi\int\rho(r)\sigma^2_r(r)r^2{\rm d}r,\quad K_t=2\pi\int\rho(r)\sigma^2_t(r)r^2{\rm d}r.
\end{equation}
In the expressions above $\rho$ is the system density, and $\sigma^2_r$ and $\sigma^2_t$ are the radial and tangential phase-space averaged square velocity components \citep[see][]{poly81,poly92}: in particular, for isotropic systems $\xi=1$. In the case of Newtonian force numerical simulations show that the ROI typically occurs for $\xi\gtrsim1.7$, but this should be considered a fiducial value, as it is also well known that the critical value of $\xi$ above which the given system is unstable depends on the specific phase-space structure of the equilibrium configuration under study \citep[see, e.g.][]{M85b,bert89,saha91,bert94,MZ97,NLC02,barnes09}.\\
\indent Due to its relevant astrophysical consequences, and its intrinsic physical interest for the understanding of collisionless systems, the ROI has been extensively studied in Newtonian gravity \citep[see e.g.][and references therein]{palmer87,Gajda}. For example, \citet{NLC02} investigated the implications of the ROI for the Fundamental Plane of elliptical galaxies, excluding the possibility that the so-called tilt of the Fundamental Plane is just due to a systematic increase of radial orbits with galaxy mass (see also \citealt{ciotti97}, \citealt{ciottilanzoni}). In a different context (\citealt{ciotti04}) it has been shown that the constraints on the maximum amount of radial anisotropy that can be sustained  by a stable stellar system can be used to dismiss some models of mass distribution in elliptical galaxies that have been obtained by using X-ray data under the assumption of hydrostatic equilibrium of the hot interstellar medium. The ROI  also plays a role in the course of Violent Relaxation (\citealt{LB67}), i.e., the collapse and virialization of cold self-gravitating systems \citep[e.g. see][]{vanAlbada82,LMS91,NLC06a}, because such collapses are dominated in their last stages by significant radial motions, and so ROI  contributes to establish the dynamical and structural properties of the quasi-relaxed final states (e.g. \citealt{M85b,bert05,sylos15}).\\
\indent From a more general point of view, it is now known that the ROI is not a specific property of Newtonian gravity.  For example \cite{NCL11} have studied the ROI in Modified Newtonian Dynamics (hereafter MOND; \citealt{milgrom83}; \citealt{BM84}), comparing anisotropic MOND systems with their Equivalent Newtonian Systems with dark matter, (i.e., systems in which the phase-space properties and total gravitational potential are the same as in the corresponding MOND systems). Numerical simulations showed that MOND systems are {\it more} prone to develop ROI than their ENSs. At the same time, MOND systems are able to support a larger amount of kinetic energy stored in radial orbits than one-component Newtonian systems with the same density distribution.\\
\indent Some natural questions then arise. What is the dependence of the ROI on the specific force law? How much the long- or short-range nature of the interparticle force influences the ROI? Are there any aspects of the ROI that are peculiar of the Newtonian force (which stands out among power-law forces for several unique mathematical properties)? In fact, though several features of the dynamical evolution of collisionless systems have found to be quite independent of the specific force law, some other aspects show a clear dependence on it. For example, in the low-acceleration regime, collisionless relaxation is much less efficient in MOND than in Newtonian gravity\footnote{ Note however that in MOND collisional relaxation seems to be faster than in Newtonian gravity. See e.g. \cite{binney04,Nipoti08}.} (\citealt{CNL07,NLC07,NLC07bis}), and a similar result is found for power-law forces with $\alpha<2$ (\citealt{dicintio2011}).\\
\indent In particular, \cite{DCN13} (see also \citealt{dicintio2011}) studied by means of direct $N-$body simulations the dissipationless collapses of cold and spherical systems of particles interacting via additive interparticle $r^{-\alpha}$ forces. The main results of these works are that, almost independently of $\alpha$, the considerably radially anisotropic final states of cold collapses have surface density profiles following the \cite{sersic68} law, and  differential energy distributions well described by an exponential over a large range of energies. At fixed initial virial ratio however, a non-monotonic trend of the S\'ersic index $m$ (the parameter determining the shape of the density profile, see e.g. \citealt{ciotti91}, \citealt{ciottibertin}) with the force index is found. Remarkably, the $\alpha=1$ case (corresponding qualitatively to the force in the deep-MOND regime), behaves very similarly to MOND, although this last theory is non-additive (the MOND field equation involves the $p$-Laplace operator).\\
\indent Prompted by these findings and by the above questions, here we set the stage for the study of ROI in the case of power-law forces, by constructing a family of radially anisotropic \cite{her} models with additive $r^{-\alpha}$ forces, and restricting to the range $1\leq\alpha<3$. Such interval contains the Newtonian $\alpha=2$ case, explores ''short-range" forces ($\alpha>2$) and also long-range forces ($\alpha<2$), down to the MOND-like case $\alpha=1$. Smaller values of $\alpha$ are excluded by this preparatory investigation (we recall that the $\alpha=-1$ case corresponds to the harmonic force, for which no relaxation takes place, see e.g. \citealt{LB82}). Power-law forces with $\alpha\geq 3$ are also excluded by the present analysis, due to their very peculiar properties (for example, self-gravitating systems with $\alpha=3$ can be virialized only for zero total energy). We notice that the study of the dynamics under the effect of power-law forces is by no means new: very important examples can be traced back to Newton's {\it Principia} (e.g. see \citealt{chandr95}).\\
\indent This paper is structured as follows. In Section 2, we construct isotropic and radially anisotropic self gravitating spherical Hernquist models with $r^{-\alpha}$ forces, and we study their phase-space consistency and the properties of their velocity dispersion profiles. In Section 3 we presents numerical results on the stability of the isotropic systems. Section 4 summarizes. 
\section{Radially anisotropic equilibria}
The first step for the realization of the $N-$body simulations of ROI for systems ruled by generalized power-law forces is the construction of their equilibrium phase-space distribution function $f$, in a framework allowing for tunable amount of radial anisotropy (i.e. different values of $\xi$). It is well known that the amount of radial orbits that can be imposed to a physically acceptable equilibrium stellar system is limited by phase-space consistency (i.e. by the positivity of the associated $f$): stability can be studied only for consistent systems. Therefore, before embarking in the study of the stability of a system it is fundamental to be reassured that its equilibrium initial condition actually exists. For these reasons in the next Section we proceed to construct the phase-space distribution function for the widely used \cite{her} profile\footnote{A Referee pointed out to us that the Hernquist (1990) profile has been found earlier as  limiting case of a generalized isochrone model (\citealt{kuzmin70,kuzmin73}, see also \citealt{O78})}, under the effect of the $r^{-\alpha}$ force.
\subsection{The phase-space distribution function}
We begin by considering the problem of the construction of the phase-space distribution function $f$ for a generic equilibrium spherical system of given density profile $\rho(r)$, under the action of the $r^{-\alpha}$ force with $1\leq\alpha<3$. In principle, different velocity anisotropy profiles can be imposed on a given density profile. In this work, in order to build systems with a tunable degree of radial anisotropy, we adopt widely used Osipkov-Merritt extension of the classical \citet{edd16} inversion: 
\begin{equation}\label{OM}
f(Q)=\frac{1}{\sqrt{8}\pi^2}\frac{\rm d}{{\rm d} Q}\int_Q^{Q_{\rm sup}}\frac{{\rm d}\rho_a}{{\rm d}\Phi}\frac{{\rm d}\Phi}{\sqrt{\Phi-Q}}=\frac{1}{\sqrt{8}\pi^2}\int_Q^{Q_{\rm sup}}\frac{{\rm d^2}\rho_a}{{\rm d}\Phi^2}\frac{{\rm d}\Phi}{\sqrt{\Phi-Q}},
\end{equation}
where for a system of finite total mass $Q_{\rm sup}=0$ for $\alpha>1$, and $Q_{\rm sup}=\infty$ for $\alpha\leq1$ (see the following discussion). Here $\Phi$ is the (self-consistent) gravitational potential, $Q=E+{J^2}/{2r_a^2}$, where $E$ and $J$ are the specific (per unit mass) particle energy and angular momentum modulus, respectively,
\begin{equation}\label{augmented}
\rho_a(r)\equiv\left(1+\frac{r^2}{r_a^2}\right)\rho(r)
\end{equation}
is the so-called {\it augmented density}, and $r_a$ is the {\it anisotropy radius} \citep[][]{O79,M85}. Due to the dependence of $f$ on global integrals of motion, it follows from the Jeans theorem that the equilibrium phase-space distribution function $f(Q)$ satisfies the Vlasov equation (also known as the Collisionless Boltzmann Equation, see e.g. \citealt{bt2008}, \citealt{bert01})
Integration of $f(Q)$ over the velocity space after multiplication by the appropriate velocity components shows that the anisotropy profile associated with eq. (\ref{OM}) is 
\begin{equation}\label{beta}
\beta(r)\equiv1-\frac{\sigma_t^2(r)}{2\sigma_r^2(r)}=\frac{r^2}{r_a^2+r^2},
\end{equation} 
where $\sigma_t(r)$ and $\sigma_r(r)$ are the tangential and radial velocity dispersion profiles of the system \citep[see e.g.][]{bt2008}. In practice, Osipkov-Merritt models are everywhere isotropic for ${r_a}\to\infty$ and become more and more radially anisotropic for decreasing values of $r_a$; at fixed $r_a$ they are isotropic for $r\ll r_a$ and radially anisotropic for $r\gg r_a$. It is easy to show that the above formulae, derived in the literature for Newtonian gravity, are independent of the specific force law adopted.\\
\begin{figure}
  \centerline{\includegraphics{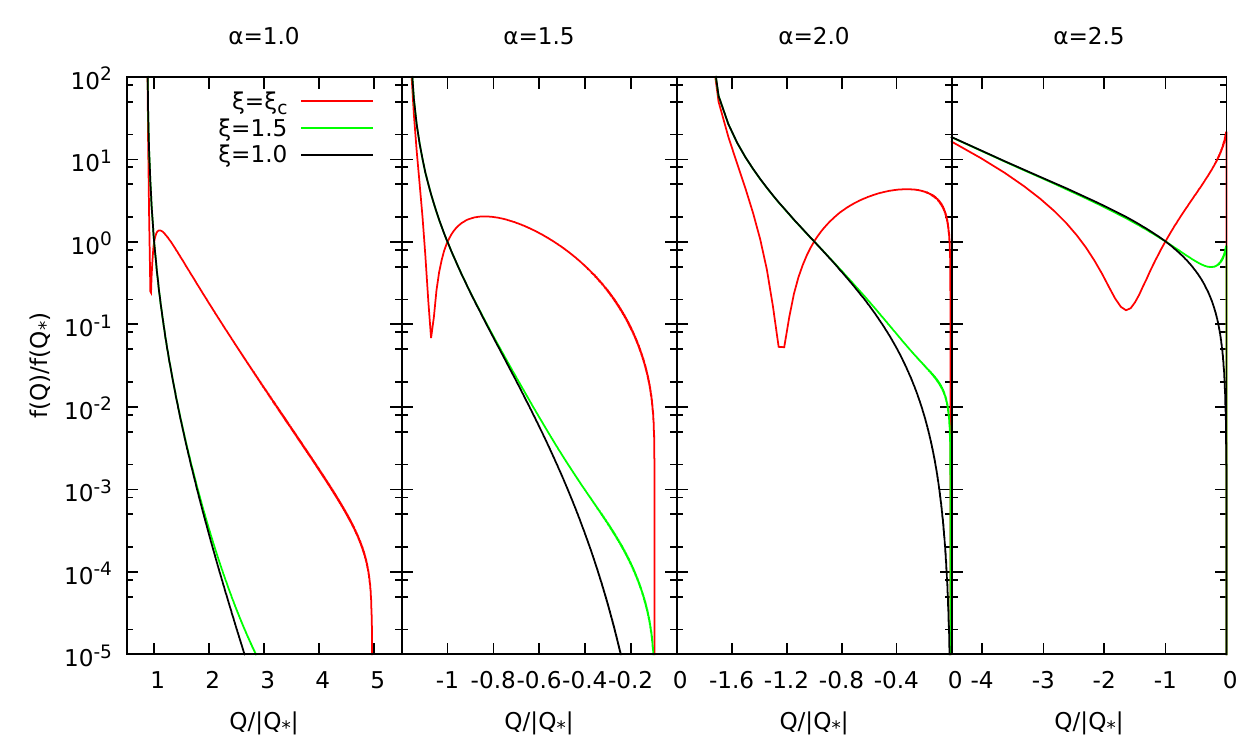}}
  \caption{Phase-space distribution functions for Osipkov-Merritt radially anisotropic Hernquist models with $\alpha=1$, 1.5, 2 and 2.5. The normalization constant is $Q_*=\Phi(r_*)$, where $r_*$ is the half-mass radius. The black curves correspond to the isotropic case ($\xi=1$), while the green ones to $\xi=1.5$, the fiducial value for the onset of ROI in Newtonian ($\alpha=2$) systems. The red curves show $f(Q)$ for values of $\xi$ near the consistency limit $\xi_c=\xi(r_{ac})$ (see Sect. \ref{sect22}).}
\label{fig1}
\end{figure}
\indent From eq. (\ref{OM}) it follows that the first step needed to recover $f$ is the knowledge of the self consistent potential $\Phi$ of the system. For a generic density $\rho(\mathbf{x})$ the $\alpha-$potential can be written as
\begin{equation}\label{pota}
\Phi(\mathbf{x})-\Phi(\mathbf{x}_0)=-\frac{G}{\alpha-1}\int_{\mathbb{R}^3}\left(\frac{1}{||\mathbf{x}-\mathbf{y}||^{\alpha-1}}-\frac{1}{||\mathbf{x}_0-\mathbf{y}||^{\alpha-1}}\right)\rho(\mathbf{y}){\rm d}^3\mathbf{y}
\end{equation}
for $\alpha\neq 1$, and as
\begin{equation}\label{potb}
\Phi(\mathbf{x})-\Phi(\mathbf{x}_0)=G\int_{\mathbb{R}^3}\ln\frac{||\mathbf{x}-\mathbf{y}||}{||\mathbf{x}_0-\mathbf{y}||}\rho(\mathbf{y}){\rm d}^3\mathbf{y}
\end{equation}
for $\alpha=1$. In the equations above $G$ is a dimensional coupling constant and $\mathbf{x}_0$ is a reference point to be chosen from convergence considerations. The value of $\Phi(\mathbf{x}_0)$ is usually fixed by arguments of convenience as illustrated below. Note that, when convergence is assured, eq. (\ref{potb}) follows from (\ref{pota}) for $\alpha\to 1$, provided that $||\mathbf{x}_0||<\infty$.\\ 
\indent Elementary integration shows that for a spherical density distribution $\rho(r)$ of {\it finite mass}, the potential can be written for $1<\alpha<3$ as
\begin{equation}\label{alfa2}
\Phi(r)=-\frac{2\pi G}{r}\int_{0}^{\infty}\frac{(r+r^\prime)^{3-\alpha}-|r-r^\prime|^{3-\alpha}}{(\alpha-1)(3-\alpha)}r^\prime\rho(r^\prime){\rm d} r^\prime,
\end{equation}
having set to zero the value of the potential for $r\to\infty$, and assumed $||\mathbf{x}_0||\to\infty$. For $\alpha=1$ some care is needed in order to define $\Phi(r)$, and it can be shown that for a system of {\it finite mass} the potential can be written as
\begin{eqnarray}\label{alfa1}
\Phi(r)=\frac{\pi G}{r}\int_{0}^{\infty}\left[(r+r^\prime)^2\ln\frac{r+r^\prime}{r_n}-(r-r^\prime)^2\ln\frac{|r-r^\prime|}{r_n}-2rr^\prime\right]r^\prime\rho(r^\prime){\rm d}r^\prime,
\end{eqnarray}
where now $\mathbf{x}_0=0$, and $r_n$ is an arbitrary normalization length: different choices of $r_n$ just correspond to a different additive constant in the value of $\Phi$.
As expected eqs. (\ref{alfa2}) and (\ref{alfa1}) are more complicated than their Newtonian counterparts, because Newton's theorems do not hold for $\alpha\neq 2$. However, in general the potential at large radii for $1<\alpha<3$ is dominated by the very simple asymptotic  monopole term
\begin{equation}\label{alfapot}
\Phi(r)\sim-\frac{GM}{(\alpha-1)r^{\alpha-1}},
\end{equation}
while in the $\alpha=1$ case the monopole term for $r\to\infty$ diverges as
\begin{equation}\label{logpot}
\Phi(r)\sim GM\ln\frac{r}{r_n}.
\end{equation}
Note that, the expansions in eqs. (\ref{alfapot}) and (\ref{logpot}) hold true also in the case of non-spherical density profiles, as can be seen from eqs. (\ref{pota}) and (\ref{potb}) for systems of finite total mass.\\
\indent In the present work we focus on systems described by the \citet{her} density distribution
\begin{equation}\label{hernquist}
\rho(r)=\frac{M}{2\pi r_c^2r(1+r/r_c)^3},
\end{equation}
where $M$ is the total mass, $r_c$ is the core radius and $r_*=(1+\sqrt{2})r_c$ is the half-mass radius, i.e. the radius of the sphere enclosing half of the total mass.\\
\indent For the distribution (\ref{hernquist}) and $1<\alpha<3$ the potential at the centre can be evaluated analytically and reads
\begin{equation}\label{pot0a}
\Phi(0)=-\frac{GM}{r_c^{\alpha-1}}\frac{2{\rm B}(3-\alpha,\alpha)}{\alpha-1},
\end{equation}
where ${\rm B}(x,y)$ is the Complete Euler Beta function. Curiously, the dimensionless factor in equation above is a non monotonic function of $\alpha$, with maximum -1 reached for $\alpha=2$, and diverging at $-\infty$ for $\alpha\to 1^+$ and $\alpha\to 3^-$. For $\alpha=1$ the central potential is instead finite and reads
\begin{equation}\label{pot01}
\Phi(0)=GM\left(1+\ln\frac{r_c}{r_n}\right);
\end{equation}
its numerical value depends on the normalization length $r_n$. For Hernquist models with $\alpha=1$ the natural choice is to adopt $r_n=r_c$. It is not surprising that the limit of eqs. (\ref{alfapot}) and (\ref{pot0a}) for $\alpha\to 1^+$ differs from eqs. (\ref{logpot}) and (\ref{pot01}), as two different choices of $\mathbf{x}_0$ have been adopted for $1<\alpha<3$ and $\alpha=1$ (see the discussion after eq. \ref{potb}). In general, the \cite{cheby} theorem on the integration of differential binomials assures that the integral in eq. (\ref{alfa2}) can be evaluated in terms of elementary functions for the Hernquist model for all rational values of $\alpha$. However, the formulae are cumbersome, so we computed eqs. (\ref{alfa2}) and (\ref{alfa1}) numerically; as a safety check we verified that the numerical results reproduce the asymptotic expansions at large radii (equations \ref{alfapot} and \ref{logpot}) as well as the values of $\Phi(0)$.
\subsection{Consistency}\label{sect22}
In Newtonian gravity $(\alpha=2)$ anisotropic Osipkov-Merritt systems of given $\rho(r)$ are characterized by a critical value of the anisotropy radius, $r_{ac}$, such that for $r_a<r_{ac}$ the corresponding models are inconsistent (\citealt{ciotti92}, see also \citealt{carollo95,ciotti96,ciotti99,CM10a,CM10b,an12}). It is therefore important for our study to determine the minimum value $r_{ac}(\alpha)$ associated with a nowhere negative (i.e. consistent) $f(Q)$. From eq. (\ref{OM}) it follows that the phase-space distribution function of an Osipkov-Merritt model (independently of the force law) can be written as
\begin{equation}
f(Q)=f_i(Q)+\frac{f_a(Q)}{r_a^2},
\end{equation}
where $f_i$ is the phase-space distribution function in the isotropic case ($r_a\to\infty$). For the Hernquist density profile we found that $f_i\geq0$ for all the explored values of $\alpha$ i.e. the isotropic Hernquist model is consistent under $r^{-\alpha}$ forces, as already known for the Newtonian case. 
\begin{figure}
  \centerline{\includegraphics{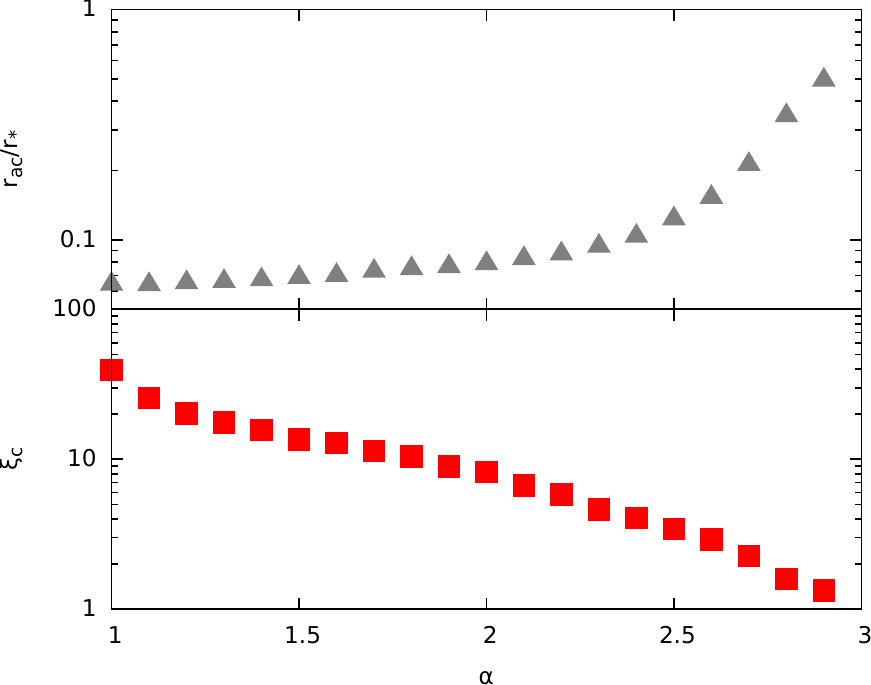}}
  \caption{Minimum value $r_{ac}$ of the anisotropy radius for phase-space consistency (in units of the half-mass radius $r_*$) of the Hernquist model with Osipkov-Merritt radial anisotropy and the corresponding value of the Fridman-Polyachenko-Shukhman index $\xi_c=\xi(r_{ac})$, as functions of the force exponent $\alpha$.}
\label{fig2}
\end{figure}
Being this assured, then the critical anisotropy radius is given by 
\begin{equation}\label{decomposition}
r_{ac}^2={\rm sup}\left[-\frac{f_a(Q)}{f_i(Q)}\right],\quad Q\in A^{-},
\end{equation}
where the sup in the r.h.s. is evaluated over the region $A^-$ of $Q$ values corresponding to $f_a<0$ \citep[see][]{ciotti00}.\\
\indent The numerically recovered phase-space distribution functions for $\alpha=1,$ 1.5, 2 and 2.5 and different values of $r_a$ (with the corresponding values of $\xi$) are presented in Fig. \ref{fig1}. Of course, $\Phi(0)\leq Q<0$ for $1<\alpha<3$, and $\Phi(0)\leq Q\leq\infty$ for $\alpha=1$. For $\xi\simeq\xi_c$, i.e. $r_a\simeq r_{ac}$ (red curves), $f(Q)$ shows a strong non-monotonicity, due to the almost dominant negative contribution of $f_a/r_a^2$ over the isotropic and positive $f_i$. A further reduction of $r_a$ would deepen even more the depression in the red curves, finally leading to a negative $f$, and to phase-space inconsistency. The relation of monotonicity of $f(Q)$ and ROI is a very important point that at this stage we cannot address, but that will be central in our forthcoming study (for example by considering the possibility to extend the Antonov laws to $r^{-\alpha}$ forces; see \citealt{bt2008}).\\ 
\indent In Fig. \ref{fig2} we show for $1\leq\alpha<3$ the values of $r_{ac}$, and their associated $\xi_c=\xi(r_{ac})$. In the Newtonian case ($\alpha=2$) $r_{ac}/r_*\simeq 0.08$, in accordance with analytic results (\citealt{ciotti96}).
From Fig. \ref{fig2} it is also apparent that systems with lower values of $\alpha$ (i.e. forces with a longer range than Newtonian) can sustain smaller values of the anisotropy radius (associated with larger values of $\xi_c$) implying that they are able to store a larger fraction of the kinetic energy in low-angular momentum orbits maintaining a nowhere negative phase-space distribution function.\\
\begin{figure}
  \centerline{\includegraphics{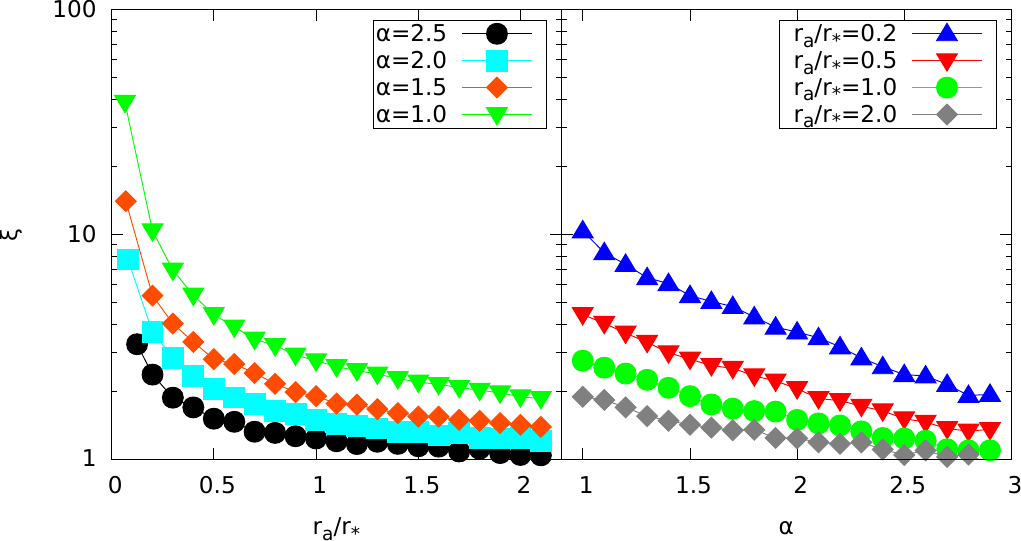}}
  \caption{Fridman-Polyachenko-Shukhman parameter $\xi$ as a function of the normalized anisotropy radius $r_a/r_*$, and of the force exponent $\alpha$, for Osipkov-Merritt radially anisotropic Hernquist models with $r^{-\alpha}$ forces.}
\label{fig3}
\end{figure}
\indent Figure \ref{fig3} (left panel) shows the parameter $\xi$ as a function of the anisotropy radius $r_a$ for $\alpha=1,$ 1.5, 2 and 2.5. As expected, $\xi$ decreases monotonically for increasing $r_a$ for all values of $\alpha$. A complementary illustration of this behaviour is given in the right panel, where the trend of $\xi$ with $\alpha$ for a selection of values of $r_a$ is given.\\ 
\indent A more detailed representation of the effects of $\alpha$ and $r_a$ on the internal dynamics of the models can be obtained by plotting the velocity dispersion profiles. The radial velocity dispersion $\sigma_r$ is given by the Jeans equation
\begin{equation}\label{jeans1}
\frac{{\rm d}\rho\sigma_r^2}{{\rm d} r}+\frac{2\beta\rho\sigma_r^2}{r}=-\rho\frac{{\rm d}\Phi}{{\rm d} r},
\end{equation}
whose solution in case of Osipkov-Merritt anisotropic systems is
\begin{equation}\label{jeans}
\rho(r)\sigma_r^2(r)=\frac{\int_r^{\infty}\rho(r^\prime)(r^{\prime2}+r_a^2)\frac{{\rm d}\Phi}{{\rm d} r^\prime}dr^\prime}{r^2+r_a^2}.
\end{equation}
It is easy to show that, for a density profile decreasing at large radii as $r^{-\lambda}$ with $\lambda>3$, and in presence of $r^{-\alpha}$ forces,
\begin{equation}\label{sigma2}
\sigma_r^2(r) \simeq \frac{GM}{r^2 + r_a^2}\left(\frac{r^{3-\alpha}}{\alpha+\lambda-3}+\frac{r_a^2 r^{1-\alpha}}{\alpha + \lambda -1}\right).
\end{equation}
In particular, for $\alpha=1$, in the limit $r\rightarrow\infty$, $\sigma_r^2$ flattens to the constant value $GM/(\lambda-2)$ due to the logarithmic nature of the potential for a system of finite mass (see eq. \ref{logpot}). Specifically, for the Hernquist density profile given by eq. (\ref{hernquist}), $\lambda=4$, and thus $\sigma_r\simeq\sqrt{GM/2}$ at large radii. In Fig. \ref{fig4} we present the radial and tangential velocity dispersion profiles $\sigma_r(r)$ and $\sigma_t(r)$ for $r_a/r_*=0.5,$ 1 and 2, obtained by numerically solving eq. (\ref{jeans}). As a further test of the numerical recovery of $f(Q)$ we verified that the profiles of $\sigma_r$ and $\sigma_t$ extracted from $N-$body realizations obtained by sampling the numerically evaluated $f(Q)$ reproduce the curves shown in Fig. \ref{fig4}. It is apparent how for increasing $\alpha$, independently of $r_a$ the components of the velocity dispersion become more and more peaked at inner radii. As expected, for the MOND-like ($\alpha=1$) case, the radial profile of $\sigma_r$ is always flat for all the explored values of $r_a$, in agreement with eq. (\ref{sigma2}) and with the numerical results of \cite{DCN13}. Once the profiles of $\sigma_r^2$ and $\sigma_t^2$ are known (see eq. \ref{beta}), it is immediate to evaluate $K_r$ and $K_t$ by numerical integration. In particular, the relatively flat profiles of $\sigma_r$ at large radii for small values of $\alpha$ are at the origin of the high values of $\xi$ with respect to those obtained for larger values of $\alpha$. The behaviour of $\xi$ as a function of $r_a$ and $\alpha$ can be also easily illustrated with the aid of a very simple toy model. In practice, one can integrate eq. (\ref{sigma2}) for a constant density ''atmosphere", $\rho(r)=\rho_0$ for $r\leq R$, under the gravitational field of a central point mass, $-GM/r^{\alpha}$. The radial velocity dispersion profile is of elementary evaluation, while $K_r$ in general is given by hypergeometric (Euler Beta) functions. However, again from the Chebyshev theorem, $K_r$ can be expressed in terms of elementary functions for all rational values of $\alpha$. $K_t$ is then obtained from the Virial theorem as $K_t=-K_r+|W|/2$ where $W$ is given in the following eq. (\ref{viriale1}), which for the toy model here considered gives $W=-4\pi G\rho_0MR^{4-\alpha}/(4-\alpha)$.
\begin{figure}
  \centerline{\includegraphics{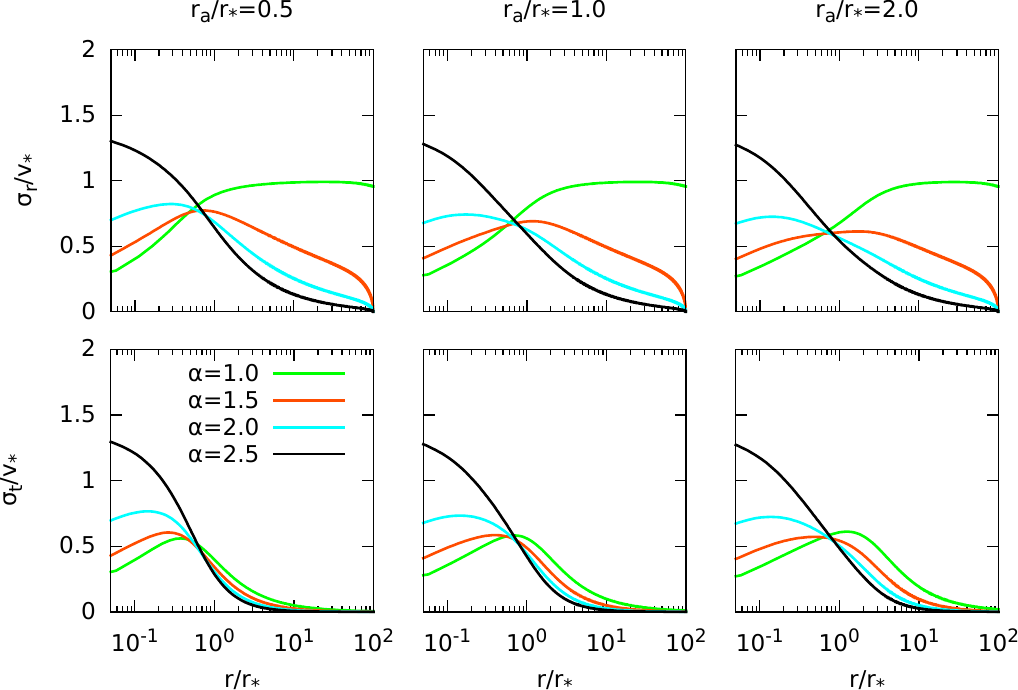}}
  \caption{Radial and tangential velocity dispersion profiles of the self-consistent radially anisotropic Hernquist model with $r^{-\alpha}$ forces, for $r_a/r_*=0.5$, 1 and 2; $r_*$ is the half-mass radius of the distribution. The velocity dispersions are expressed in units of $v_*=r_*/t_*$, and $t_*$ is given in eq. (\ref{tdyn}). The flat outer parts of the green curve ($\alpha=1$) in the top panels are in  excellent agreement with the expected value from eq. (\ref{sigma2}), and $\lambda=4$.}
\label{fig4}
\end{figure}
\section{Stability of isotropic systems}
\begin{figure}
  \centerline{\includegraphics{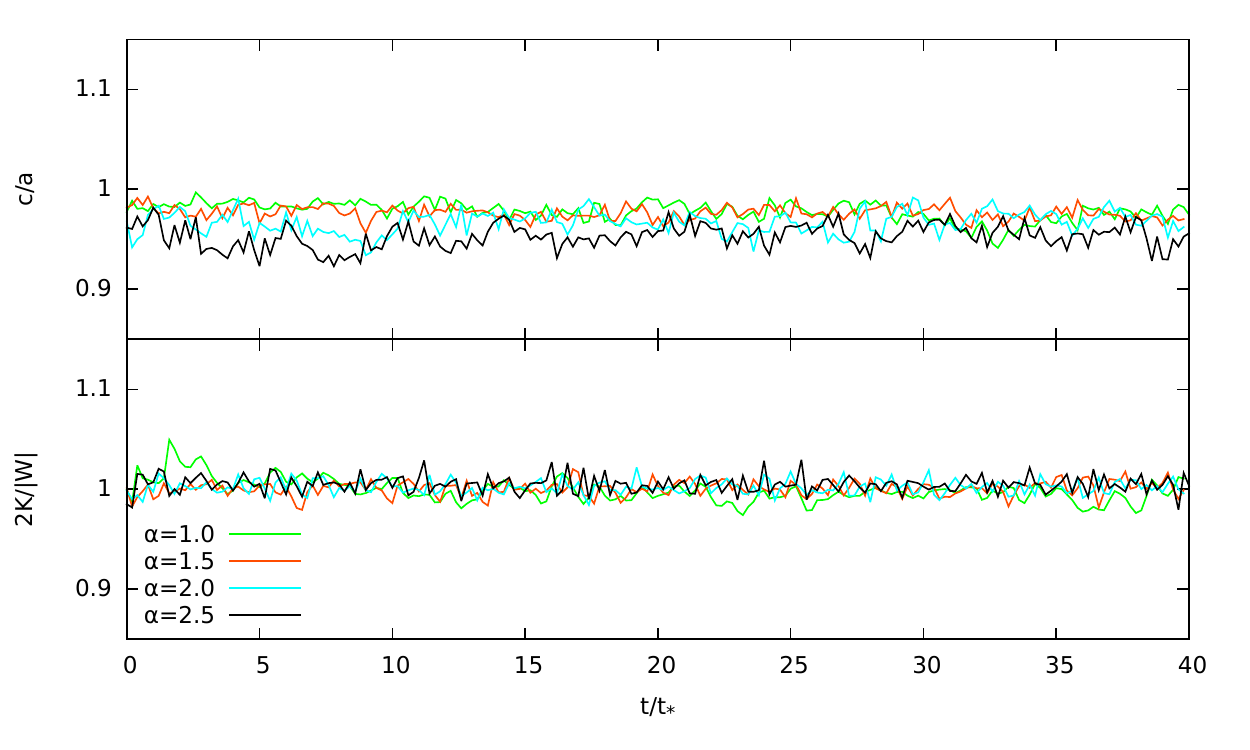}}
  \caption{Evolution of the fiducial axial ratio $c/a$ and the virial ratio $2K/|W|$ for $N-$body simulations of isotropic Hernquist models with $r^{-\alpha}$ forces, and $\alpha=1$, 1.5, 2, 2.5.}
\label{fig5}
\end{figure}

As a first sample of the numerical simulations that will be performed in a forthcoming work, we present here direct $N-$body simulations aimed at testing the stability of isotropic ($\xi=1$) Hernquist models for various values of $\alpha$. The initial conditions are generated as in \cite{NLC02} with a Monte Carlo sampling of $f(Q)$. The equations of motion for $N=25000$ equal-mass particles are integrated with a second order symplectic scheme with fixed timestep $\Delta t=t_*/100$, where the natural dynamical time-scale is defined as
\begin{equation}\label{tdyn}
   t_{*}\equiv\sqrt{\frac{2r_*^{\alpha+1}}{GM}}
\end{equation}
(see \citealt{dicintio2011} and \citealt{DCN13}). In the code, the divergence of force and potential at vanishing inter-particle separation is cured with the introduction of a softening length $\epsilon$, so that in practice the potential due to each particle is smoothed as $\Phi\propto 1/(r^2+\epsilon^2)^{(\alpha-1)/2}$ for $\alpha>1$, and $\Phi\propto \ln(\sqrt{r^2+\epsilon^2})$ for $\alpha=1$. The optimal $\epsilon$ (in units of $r_*$) is fixed so that the softened force on a particle placed at $\simeq5r_*$ from the centre of mass of the system differs less than $0.01\%$ from the unsoftened force. Therefore the softening length $\epsilon$ is different for different $\alpha$: $\epsilon=0.05,$ 0.07, 0.1 and 0.125 for $\alpha=1$, 1.5, 2 and 2.5 respectively.\\
\indent The absence of the analogue of the Poisson equation\footnote{More specifically $r^{-\alpha}$ forces do not obey a Poisson-like equation, belonging to the class of the so-called Riesz potentials, associated to fractional Laplace operators (\citealt{stein70}).} for $r^{-\alpha}$ forces and $\alpha\neq 2$,  points at direct $N-$body simulations as the most natural approach for numerical studies. However, it is also important to note the well known property that potentials associated with the $r^{-\alpha}$ force can be expanded in terms of Gegenbauer polynomials for $\alpha\neq 1$,  and of Chebyshev polynomials for $\alpha=1$ (\citealt{arfken}), thus opening the way to the construction of tree-codes (see e.g., \citealt{srinivasan}).\\  
\indent In the upper panel of Fig. \ref{fig5} we show, for some representative value of $\alpha$, the time evolution of the minimum-to-maximum semiaxis ratio $c/a$ of the systems. The semiaxes are defined from the eigenvalues $I_1\geq I_2\geq I_3$ of the tensor $I_{ij}=\sum_kr_i^{(k)}r_j^{(k)}$ calculated within the radius containing 0.85\% of the total mass $M$ (\citealt{MZ97,NLC02}).
In the lower panel of the same figure we also show the 
time evolution of the corresponding virial ratio $2K/|W|$, where $K$ is the total kinetic energy and
\begin{eqnarray}\label{viriale1}
W=-\int \rho(\mathbf{x})\langle\mathbf{x},\nabla\Phi\rangle d^3\mathbf{x}=\sum_{j\neq i=1}^N m_i\langle\mathbf{x}_i,\mathbf{a}_{ji}\rangle
\end{eqnarray}
is the virial function. The second expression holds for a discrete system of particles with masses $m_i$, where 
\begin{equation}
\mathbf{a}_{ji}=-Gm_j\frac{\mathbf{x}_i-\mathbf{x}_j}{||\mathbf{x}_i-\mathbf{x}_j||^{\alpha+1}}
\end{equation}
is the acceleration on particle $i$ due to particle $j$. We recall that for $\alpha\neq 1$ $W$ is related to the system total potential energy $U$ by the identity
\begin{equation}
W=(\alpha-1)U,
\end{equation}
where
\begin{equation}
U=\frac{1}{2}\int\rho(\mathbf{x})\Phi(\mathbf{x})d^3\mathbf{x}=\frac{1}{2}\sum_{i\neq j=1}^N m_i\Phi_{ji},
\end{equation}
and in the case of a discrete system $\Phi_{ji}$ is the potential on particle $i$ due to particle $j$. 
For the $\alpha=1$ case (as for systems in deep-MOND regime) the virial function is independent of time, being: 
\begin{equation}
W=-\frac{GM^2}{2} 
\end{equation}
for a continuous system, and
\begin{equation}
W=-\frac{G}{2}\sum_{i\neq j=1}^n m_im_j
\end{equation}
for a discrete system (see \citealt{DCN13}).\\
\indent From Fig. \ref{fig5} it is apparent that the spherical symmetry as well as the equilibrium of the initial conditions is preserved for all the considered values of $\alpha$ up to $t/t_{*}=40$, when the simulations end. 
\begin{figure}
  \centerline{\includegraphics{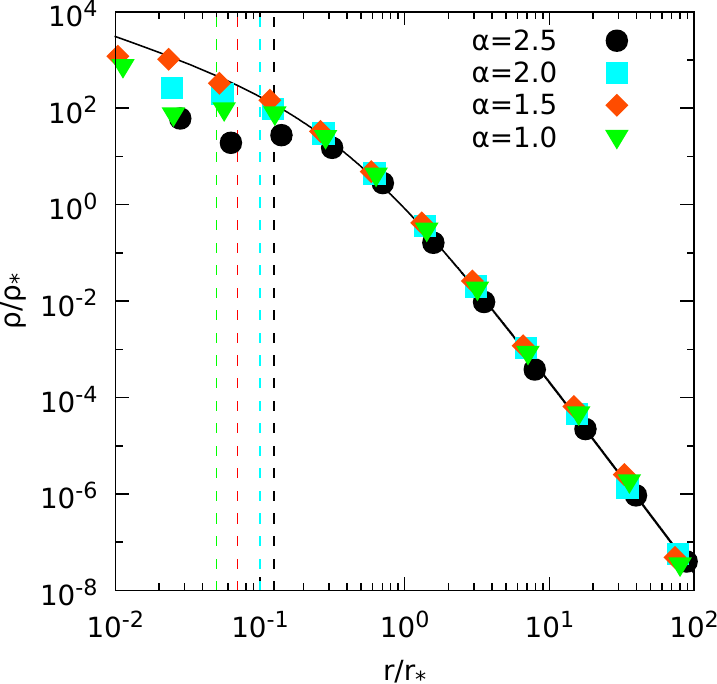}}
  \caption{Final ($t/t_*=40$) density profiles of the same systems as in Fig. \ref{fig5}. Here $\rho_*=\rho(r_*)$. The black solid line is the initial density profile given by eq. (\ref{hernquist}). The vertical dashed lines indicate the softening length $\epsilon$ used in the simulations.}
\label{fig6}
\end{figure}
The fluctuations are of the order of $\simeq5\%$ in $c/a$ and $\simeq1\%$ in $2K/|W|$. The stability of the isotropic systems is confirmed also by Fig. \ref{fig6} where we compare the final density profile with the initial profile given by eq. (\ref{hernquist}). The initial density profile is preserved down to $r/r_*\simeq0.025$ in the best case ($\alpha=1.5$) and to $r/r_*\simeq0.8$ in the worst case ($\alpha=2.5$). In all cases the final density profiles are indistinguishable from the initial profiles for $r\gtrsim1.4\epsilon$. In the forthcoming study we will perform a quantitative monitoring of the deterioration of the central profile as a function of $\alpha,\epsilon$ and $N$, by considering the time evolution of the Lagrangian radii containing a fixed fraction of the total mass.
\section{Discussion and conclusions}
In this preparatory exploration of the ROI in spherically symmetric collisionless equilibrium systems with $r^{-\alpha}$ forces we limited ourselves to a first important step, i.e. the construction of Osipkov-Merritt radially anisotropic self-consistent equilibrium models. We have recovered the potential generated by the Hernquist density distribution under $r^{-\alpha}$ forces and we have calculated the corresponding phase-space distribution functions for isotropic and anisotropic Osipkov-Merritt models in the range of values $1\leq\alpha<3$. We numerically determined, as a function of $\alpha$, the minimum value of the anisotropy radius for consistency $r_{ac}$ and the associated critical value of the Fridman-Polyachenko-Shukhman index $\xi_c$, and we computed the radial and tangential velocity dispersion profiles for a few representative consistent systems.  The main results of this paper can be summarized as follows:
\begin{enumerate}
\item  For $1\leq\alpha<3$ the distribution functions of the isotropic $r^{-\alpha}$ Hernquist models are monotonically decreasing and everywhere positive. Their $N-$body realizations are numerically stable up to $\simeq40t_*$.
\item For anisotropic Hernquist models the minimum value of the anisotropy radius for consistency $r_{ac}$ increases with increasing $\alpha$, and the corresponding $\xi_c$ decreases. Therefore, systems with lower $\alpha$ can be generated with a higher amount of radial anisotropy. This appears to be in agreement with the results of \cite{DCN13}, who find increasingly anisotropic end products of dissipationelss collapses, when reducing $\alpha$ from 3 to 1.
\item As general trend, for all values of $\alpha$ the Fridman-Polyachenko-Shukhman index $\xi$ decreases for increasing $r_a$, and increases with decreasing $\alpha$ at fixed $r_a$.\\ 
\end{enumerate}
In the natural follow-up of this preparatory work we will determine, as a function of $\alpha$, the critical $\xi$ for the onset of ROI, in order to understand how much this process depends on the short or long range nature of the interaction law and we will study the structural and dynamical properties of the final systems.
\section*{Acknowledgements}
The two Referees are acknowledged for important comments that improved the presentation. L.C. and C.N. acknowledge financial support from PRIN MIUR 2010-2011, project ''The Chemical and Dynamical evolution of the Milky Way and the Local Group Galaxies", prot. 2010LY5N2T. 
\bibliographystyle{jpp}
\bibliography{roi}
\end{document}